\documentclass[preprint,prb,showpacs,preprintnumbers,amsmath,amssymb,superscriptaddress]{revtex4}

\usepackage{graphicx}
\usepackage{dcolumn}
\usepackage{bm}

\begin{document}


\title{Nonlocality-controlled interaction of spatial solitons in nematic liquid crystals }


\author{Wei Hu}
\affiliation{Laboratory of Photonic Information Technology, South
China Normal University, Guangzhou 510631, P. R. China}
\affiliation{Laboratory of Light Transmission Optics, South China
Normal University, Guangzhou 510631, P. R. China}
\author{Tao Zhang}
\affiliation{Laboratory of Photonic Information Technology, South
China Normal University, Guangzhou 510631, P. R. China}
\affiliation{Laboratory of Light Transmission Optics, South China
Normal University,  Guangzhou 510631, P. R. China}
\author{Qi Guo}
\email[]{guoq@scnu.edu.cn} \affiliation{Laboratory of Photonic
Information Technology, South China Normal University, Guangzhou
510631, P. R. China} \affiliation{Laboratory of Light Transmission
Optics, South China Normal University,  Guangzhou 510631, P. R.
China}
\author{Li Xuan}
\affiliation{State Key Lab of Applied Optics, Changchun Institute of
Optics, Fine Mechanics and Physics, Chinese Academy of Sciences,
Changchun, 130033, P. R. China}
\author{Sheng Lan}
\affiliation{Laboratory of Photonic Information Technology, South
China Normal University, Guangzhou 510631, P. R. China}



\date{\today}

\begin{abstract}
We demonstrate experimentally that interaction between nonlocal
solitons in nematic liquid crystals (NLC) can be controlled by the
degree of nonlocality. For a given beam width, the degree of
nonlocality can be modulated by changing the pretilt angle
$\theta_0$ of NLC molecules through bias voltage $V$. As $V$
increases (so does $\theta_0$), the degree of nonlocality decreases.
When the degree of nonlocality is below a critical value, the
solitons behave in the way like their local counterpart, i.e.,
in-phase solitons attract while out-of-phase solitons repulse each
other. Such a voltage-controlled interaction between the solitons
can be readily implemented in experiments.
\end{abstract}


\maketitle



The interaction properties of two spatial optical solitons depend on
the phase difference between them, their
coherence,\cite{Stegeman1999,TSKu2005} and the nonlinear nonlocality
of the materials in which the solitons propagate. For a local
Kerr-type nonlinearity, two coherent bright solitons attract (or
repulse) each other when they are in-phase (or out-of-phase).
On the other hand, if the solitons are mutually
incoherent\cite{Anderson-pre-85} or the nonlinear nonlocality of the
materials is strong enough,\cite{Snyder1997,OL2002,xie-oqe-05} the
soliton interaction is always attractive, independent of the phase
difference. Very recently, Ku et al.\cite{TSKu2005} demonstrated
both theoretically and experimentally that the interaction of the
solitons can be controlled by varying their total coherence. Thus,
the two out-of-phase solitons can repulse or attract each other,
depending on whether the coherence parameter is below or above a
threshold. Similar dependence of the interaction behavior on
nonlocality was also theoretically predicted by Rasmussen et
al.\cite{Rasmussen2005} based on the (1+1)-Dimensional mode of the
nematic liquid crystal (NLC), i.e., there exists a critical degree
of nonlocality above which the two out-of-phase solitons will
attract each other.

The NLC with a pretilt angle induced by an external low-frequency
electric field has been confirmed\cite{PRL2003,PRL2004} to be a
typical material with the strongly nonlocal (referred also as highly
nonlocal in some papers) nonlinearity.\cite{Snyder1997}
%
%
In the previous works concerning the single soliton
propagation\cite{PRL2003,PRL2004,APL2000} and the soliton
interaction\cite{Rasmussen2005,OL2002,Fratalocchi-mclc-2004} in the
NLC, however, the peak of the pretilt angle was always made to be
$\pi/4$ in order to maximize the nonlinearity. As will be seen
later, the degree of nonlocality can only be modulated by changing
the beam width in this case, which
is not convenient in practice. Recently, Peccianti et al. have shown
that\cite{OL2005} the nonlocality can be varied by changing the
pretilt angle via a bias voltage. In this letter, we use the (1+2)-D
model with an arbitrary pretilt angle\cite{OL2005} to describe the
(1+2)-D soliton interaction in the NLC. By defining a general
characteristic length of the nonlinear nonlocality for the NLC,
a voltage-controlled degree of nonlocality is shown to be achieved
conveniently. In experiments, we observe the
nonlocality-controllable (through the change of the bias voltage)
transition from attraction to repulsion of the two out-of-phase
solitons in the NLC.


Let us consider the (1+2)-D model of light propagation in a cell
containing the NLC. The configuration of the cell and the coordinate
system are the same as in the previous
works.\cite{PRL2003,PRL2004,APL2000,OL2005,Fratalocchi-mclc-2004,peccianti-jnopm-o3}
The optical field polarized in $x$-axis with envelope $A$ propagates
in $z$-direction. An external low-frequency electric field $E_{RF}$
is applied in $x$-direction to control the initial tilt angle of the
NLC.
The evolution of the paraxial beam $A$ and the tilt angle $\theta$
can be described by the system\cite{peccianti-jnopm-o3,PRL2003}
\begin{eqnarray}
    2i k\partial_z A +\nabla_\perp^2 A + k_0^2 \epsilon_a^{op} \sin(\theta +
    \theta_0) \sin(\theta-\theta_0) A = 0,\label{nlse0}
\\
    2K \nabla_\perp^2\theta +\epsilon_0 \left(\epsilon_a^{RF} E_{RF}^2+
\epsilon_a^{op}
    \frac{|A|^2}{2}\right)\sin(2\theta) = 0, \label{roe0}
\end{eqnarray}
where $\theta_0$ is the peak-tilt of the NLC molecules in the
absence of the light,
$K$ is the average elastic constant of the NLC, $\nabla_\perp^2
=\partial_x^2 +\partial_y^2$, $k^2=k_0^2(n_\perp^2+
\epsilon_a^{op}\sin^2 \theta_0)$, $\epsilon_a^{op}=
n_\parallel^2-n_\perp^2$ and $\epsilon_a^{RF}=
\epsilon_0(\epsilon_\parallel-\epsilon_\perp$) characterize
respectively the optical and low-frequency dielectric anisotropies
with respect to the major axis of the NLC molecules (director). In
Eq.~(\ref{roe0}), the term $\partial_{z}^2\theta$ has been canceled
out because the dependence of $\theta$ on $z$ is proven to be
negligible.\cite{Rasmussen2005,peccianti-jnopm-o3} In the absence of
the laser beam, the pretilt angle $\hat{\theta}$ depends only on
$x$\cite{peccianti-jnopm-o3}
\begin{equation}
    2K \partial_x^2 \hat{\theta} + \epsilon_0 \epsilon^{RF}_a E_{RF}^2
    \sin(2\hat{\theta}) = 0. \label{pretilt}
\end{equation}

Furthermore, the system that describes $A$ and the optically induced angle
perturbation $\Psi$ [$\theta =\hat{\theta}+(\hat{\theta}/\theta_0)\Psi$] can be
simplified as\cite{peccianti-jnopm-o3,OL2005}
\begin{eqnarray}
    2i k\partial_z A +\nabla_\perp^2 A + k_0^2 \epsilon_a^{op}
    \sin(2 \theta_0) \Psi A = 0,\label{nlse1}
    \\
    \nabla_\perp^2\Psi -
\frac{1}{w_m^2} \Psi
    + \frac{\epsilon_0\epsilon_a^{op}}{4K}\sin(2\theta_0)|A|^2 = 0, \label{roe1}
\end{eqnarray}where a parameter $w_m>0$ for $|\theta_0|\leq \pi/2$, which
reads
\begin{equation}
w_m(\theta_0) =\frac{1} {E_{RF}(\theta_0)}\left\{\frac{2\theta_0 K}
{\epsilon^{RF}_a \sin(2\theta_0)\left[1- 2 \theta_0 \cot(2\theta_0)
\right]}\right\}^{1/2}.\label{wm}
\end{equation}

Introducing the normalization that $X=x/w_0$, $Y=y/w_0$,
$Z=z/(2kw_0^2)$, $a=A/A_0$, and $\psi=\Psi/\Psi_0$, where $A_0=
4\sqrt{\pi K/\epsilon_0}/k_0\epsilon_a^{op} w_0^2$, $\Psi_0=
\sin(2\theta_0)/k_0^2w_0^2\epsilon_a^{op}$, and $w_0$ the initial
beam width,
we have the dimensionless system
\begin{eqnarray}
    i\partial_Z a + \nabla_{XY}^2 a +\gamma \psi a=0, \label{nlse2}
    \\
    \nabla_{XY}^2\psi -\alpha^2 \psi +4\pi|a|^2=0, \label{roe2}
\end{eqnarray}
where $\nabla_{XY}^2 =\partial_X^2 +\partial_Y^2$, and
\begin{equation}\label{alpha}
    \gamma=\sin^2(2\theta_0), ~~\alpha=w_0/w_m.
\end{equation}
For a symmetrical geometry,\cite{explain-1} Eq.~(\ref{roe2}) has a
particular solution $\psi(x,y) =(4\pi/\alpha^2)\int
R(x-x^\prime,y-y^\prime)|a(x^\prime,y^\prime)|^2 d x^\prime d
y^\prime$, and $R(x,y) =(\alpha^2/2\pi)K_0(\alpha\sqrt{x^2+y^2})$,
where $K_0$ is the zero-th order modified Bessel function.

\begin{figure}
\includegraphics[width=8cm]{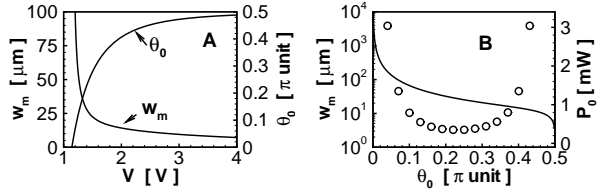}
\caption{(a) The characteristic length $w_m$ and the pretilt angle
$\theta_0$ of the NLC versus the bias voltage $V$. (b) The
characteristic length $w_m$ (a solid curve) and the critical power
of a single soliton (circles) versus the pretilt angle $\theta_0$.
The parameters are for a $80\mu m$-thick cell filled with the NLC
(TEB30A).\cite{explain-4} \label{1}}
\end{figure}

We define $w_m$ in Eq.~(\ref{wm}) as the general characteristic
length of the nonlinear nonlocality for the NLC,\cite{explain-2}
then it is obvious that the factor $\alpha$ in Eq.~(\ref{alpha})
indicates the degree of nonlocality, as defined in
Ref.\onlinecite{cao-wlxb-05} for the specific case of Gaussion
response function. A monotonous function of $\theta_0$ on $E_{RF}$
is given by Eq.~(\ref{pretilt}), and it can be approximated
as\cite{OL2005}
$\theta_0\approx(\pi/2)[1-(E_{FR}/E_{RF})^3]$
when $E_{RF}$ is higher than the Fre$\grave{e}$derichsz threshold
$E_{FR}$. Therefore, we can clearly observe from Eq.~(\ref{wm}) that
$w_m$ is determined only by $E_{RF}$ (or by the bias $V$), or
equivalently by the peak-pretilt angle $\theta_0$ for a given NLC
cell configuration, as shown in Fig.\ref{1}.
When the bias is properly chosen so that
$\theta_0=\pi/4$,\cite{PRL2003} $w_m$ is fixed and $\alpha$ can be
modulated only by changing $w_0$. This is the case discussed in
Ref.\onlinecite{Rasmussen2005}. For a given beam width, however,
$\alpha$ can be changed continuously by $w_m$ through continuously
varying the bias. As a result, the voltage-controlled degree of
nonlocality through the medium of the pretilt $\theta_0$ of the NLC
can be achieved conveniently.
With the decrease of the bias, $\theta_0$ goes from $\pi/2$ to 0,
then $\alpha$ varies from $\infty$ to 0 for a fixed $w_0$ and the
degree of nonlocality increases from locality to strong nonlocality.
In addition, we can see the factor $\gamma$ in Eq.~(\ref{alpha})
stands for the nonlinear couple between $A$ and $\Psi$. It reaches
maximum when $\theta_0=\pi/4$, which takes major responsibility for
the lowest critical power\cite{Snyder1997,cao-wlxb-05} of a single
soliton nearby $\theta_0=\pi/4$, as shown in Fig.\ref{1}(b). As the
pretilt angle approaches $\pi/2$, the critical power increases
sharply, while $w_m$ decreases to zero and $\alpha$ increases to
$\infty$ for the given $w_0$ (the degree of nonlocality decreases,
moving towards locality).

\begin{figure}
\includegraphics[width=8cm]{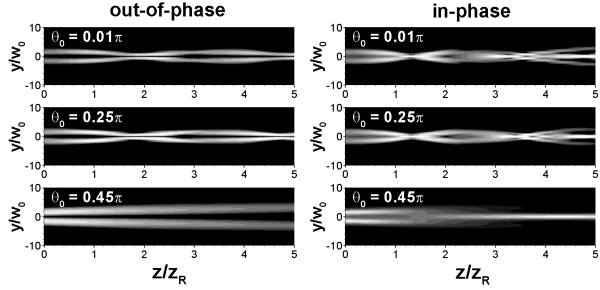}
\caption{Numerical simulation results of the interactions of the
in-phase and the out-of-phase solitons.
The width for each soliton is $4\mu$m, and the input power is
$1.1m$W. The separation and the relative angle between two solitons
are respectively $12\mu$m and $0.57^o$($\tan 0.57^o =0.01$).
\label{2}}
\end{figure}

To show the influence of the pretilt angle $\theta_0$ (or
equivalently the degree of nonlocality for a fixed $w_0$) on the
interaction between the two solitons, we have carried out numerical
simulations directly based on Eqs.~(\ref{nlse0})-(\ref{pretilt}).
The simulation results show that there exist critical values for the
degree of nonlocality below (or above) which two out-of-phase
solitons will repulse (or attract) each other. The critical values
depend on the initial separation and relative angle in the
($y,z$)-plane between the solitons. These results agree with the
prediction based on the (1+1)-D model.\cite{Rasmussen2005} The
critical degree of nonlocality for the two parallel solitons is very
weak so that the corresponding critical pretilt angle $\theta_{0c}$
is very close to $\pi/2$, leading to a very high critical power for
the soliton state [see Fig.~\ref{1}(b)].
However, the use of a relative angle will significantly increase the
critical degree of nonlocality and make $\theta_{0c}$ not be close
to $\pi/2$. Hence, the critical powers for different pretilt angles
around $\theta_{0c}$ do not differentiate too much. This makes it
possible to observe the soliton states at a fixed input power for
different pretilt angles (or different bias
voltages).\cite{explain-3}
Figure~\ref{2} presents the simulation results of two solitons with
a relative angle of $0.57^o$ for different values of $\theta_0$. We
can see that for $\theta_0 \leq \pi/4$, the nonlocality is strong
enough to guarantee the attraction of both the in-phase and the
out-of-phase solitons. However, the degree of nonlocality becomes
lower than the critical degree of nonlocality when
$\theta_0=0.45\pi$. In this case, the out-of-phase solitons begin to
repulse each other and the in-phase solitons remain attraction.

\begin{figure}
\includegraphics[width=8cm]{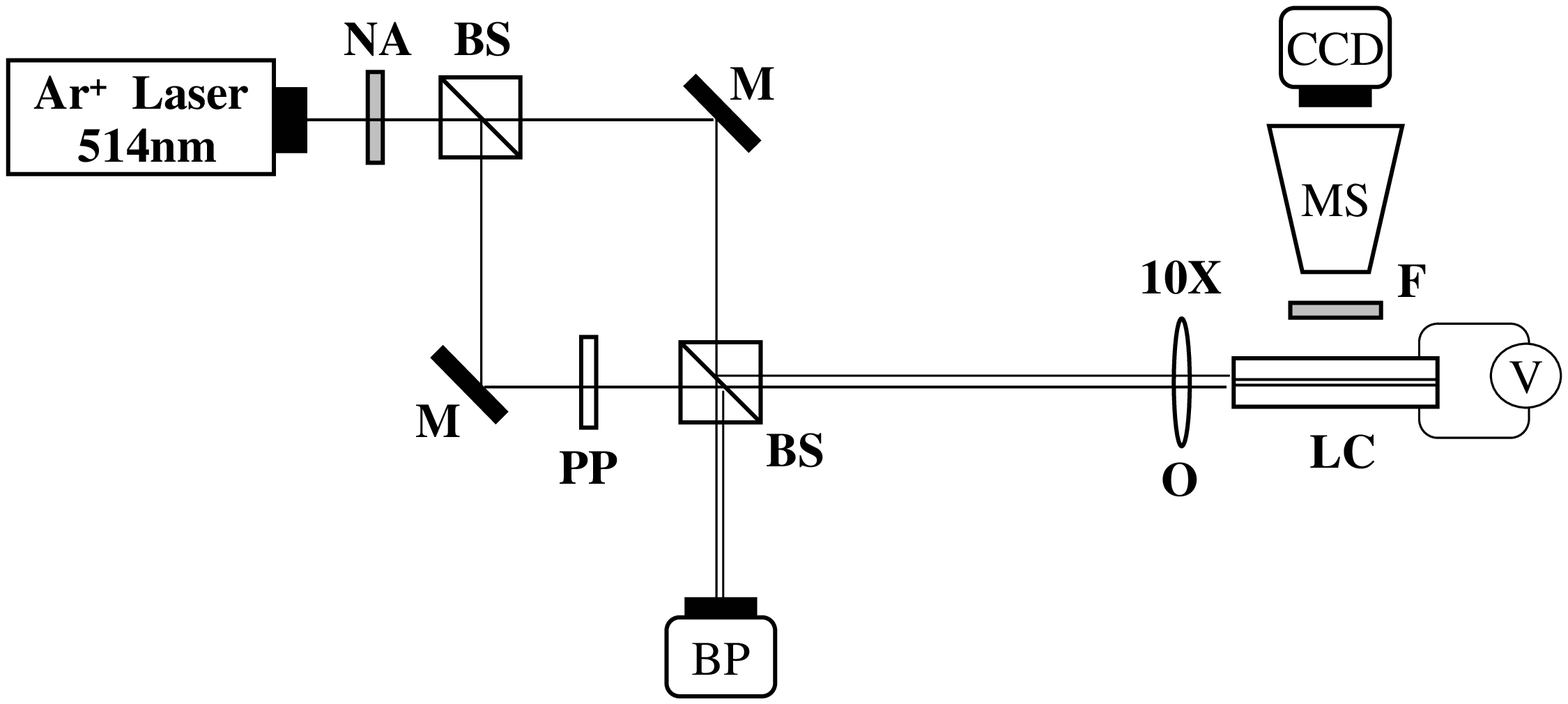}
\caption{Scheme of the experimental setup. NA, neutral attenuator;
BS, beam splitters; M, plate mirror; PP, parallel-face plate for
adjusting the phase difference; O, 10$\times$ microscope objective;
LC, liquid crystal cell; MS, microscope; F, laser-line filter; BP,
beam profiler. \label{3}}
\end{figure}

\begin{figure}
\includegraphics[width=10cm]{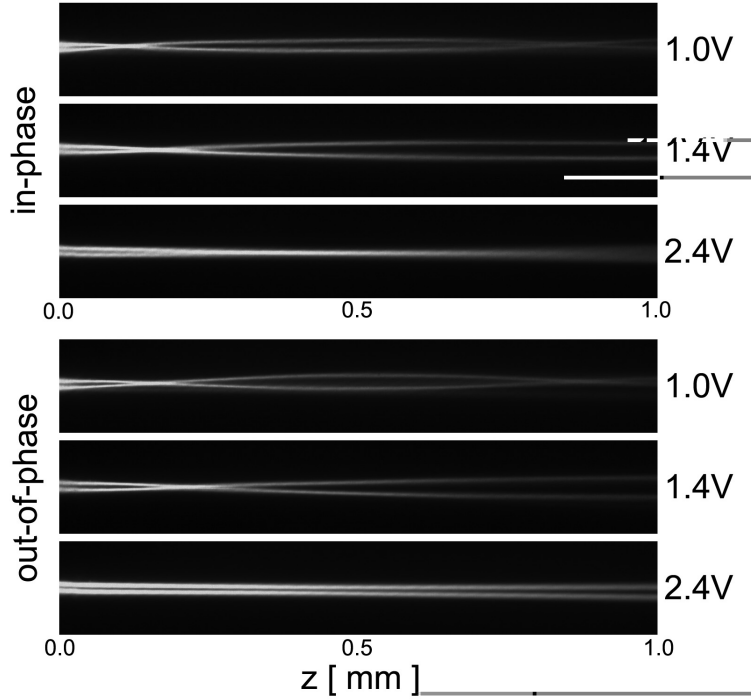}
\caption{Photos of the beam traces for the soliton pair propagation
in the NLC cell.
The biases applied on the LC are $1.0$V, $1.4$V and $2.4$V,
corresponding to the pretilt angles of $0.01\pi$, $0.25\pi$, and
$0.45\pi$, respectively. \label{4}}
\end{figure}

The experimental setup is illustrated in Fig.~\ref{3}. The laser
beam from an argon-ion laser is split into two beams, then they are
combined together with a small separation through the other
beam-splitter and launched into a $80\mu m$-thick NLC cell by a
10$\times$ microscope objective. The beam width at the focus $w_0$,
the separation $d_s$ and relative angle $\beta$ between the two
beams are measured by an edged-scanning beam profiler when the NLC
cell is removed. The phase difference between the two beams is
adjusted by the rotation of a $1.8mm$-thick parallel-face plate, and
measured through the interference pattern by the beam profiler
located on the other branch after the second beam-splitter. The cell
is filled with the NLC TEB30A (from SLICHEM China Ltd.), whose
$n_\parallel = 1.6924$, $n_\perp = 1.5221$, $K\approx 10^{-11}N$,
$\epsilon_a^{op}=0.5474$, and $\epsilon_a^{RF}=9.4\epsilon_0$. The
Fre$\grave{e}$dericksz threshold $V_t \approx 1.14$V for the $80\mu
m$-thick cell.
The launched power for each beam is fixed to 7mW when the bias is
changed, and the other parameters for the beams in the NLC are
$w_0=3.2\mu$m, $d_s=10\mu$m, and $\tan\beta=0.011$. When the phase
difference is adjusted to $0$ or $\pi$, we record the beam traces
for the different biases by the CCD camera, as shown in Fig.\ref{4}.

Let us compare the photos for the in-phase and the out-of-phase
solitons when the bias $V=1.4$V ($\theta_0 \approx \pi/4$). They are
almost the same for both cases. It means for $\theta_0 = \pi/4$ the
degree of nonlocality is strong enough to eliminate the dependence
of the interactions on the phase difference between the solitons. In
this case, $w_m \approx 25.3\mu$m, which is bigger than the
separation of the two beams, and $\alpha =0.126$ for the
$3.2\mu$m-width solitons.

For the bias $V=1.0$V sightly lower than the threshold $V_t = 1.14$V
(a small tilt angle in the sample educes some reorientation at the
bias voltage lower than the threshold\cite{OL2005}), the pretilt
angle $\theta_0$ is nearly zero and nonlocality is much stronger
than that when $V=1.4$V. For this reason, a second cross point is
observed for both the in-phase and the out-of-phase solitons.

When the bias $V$(pretilt angle $\theta_0$) increases, the degree of
nonlocality $1/\alpha$ and the characteristic length $w_m$ decrease.
For $V=2.4$V ($\theta_0 \approx 0.45\pi$), we have $w_m \approx
11\mu$m, which approximately equals to the separation between the
two solitons. In this case, we observe the attraction of the
in-phase solitons and the repulsion of the out-of-phase solitons. We
also see the two in-phase solitons fused into one soliton, which is
qualitatively same with the numerical simulation result in
Fig.~\ref{2}.


 In conclusion, we have investigated theoretically and experimentally the interactions of the nonlocal spatial
 solitons in the NLC when the applied bias is adjusted. Given is a general
 definition of the characteristic length of the nonlinear
 nonlocality for the NLC, which is the function of the bias through the medium of the pretilt angle. Hence,
 the voltage-controllable
degree of nonlocality in the NLC can be implemented expediently. We
experimentally observe the transition from attraction to repulsion
of the two out-of-phase solitons in the NLC as the degree of
nonlocality decreases via increasing the bias. Such a
voltage-controllable soliton interaction might have its potential
applications in developing all-optical signal processing devices.

This work was supported by the Natural Science Foundation of
Guangdong Province of China (Grant No. 04105804), and partially
supported by the National Natural Science Foundation of China (Grant
Nos. 60278013, 10474023, and 60277033).


\begin{thebibliography}{Reference}
\bibitem{Stegeman1999}About the dependence on the
phase-difference and the coherence, see, G. I. Stegeman and M.
Segev,
Science {\bf 286}, 1518 (1999) and references therein.

\bibitem{TSKu2005} T.S. Ku, M.-F. Shih, A. A. Sukhorukov, and Y. S. Kivshar,
\prl {\bf 94}, 063904 (2005).

\bibitem{Anderson-pre-85}D. Anderson and M. Lisak, \pra {\bf32}, 2270 (1985).

\bibitem{Snyder1997}A. W. Snyder and D. J. Mitchell,
Science {\bf 276}, 1538 (1997).

\bibitem{OL2002}M. Peccianti, K. Brzdakiewicz, and G.
Assanto,
\ol 27, 1460 (2002).

\bibitem{xie-oqe-05}Y. Xie and Q. Guo, Opt. Quant. Electron. {\bf36},
1335 (2004).


\bibitem{Rasmussen2005}P. D. Rasmussen, O. Bang, and Wieslaw Kr$\acute{o}$likowski,
\pre {\bf 72}, 066611(2005).


\bibitem{PRL2003}C. Conti, M. Peccianti, and G. Assanto,  \prl {\bf 91}, 073901(2003).


\bibitem{PRL2004}C. Conti, M. Peccianti, and G. Assanto, \prl {\bf 94}, 113902(2004).

\bibitem{APL2000}M. Peccianti, A. De Rossi, G. Assanto, A. De Luca, C. Umeton, and I. C. Khoo,
\apl {\bf 77}, 7(2000).

\bibitem{Fratalocchi-mclc-2004}A. Fratalocchi, M. Peccianti, C. Conti, and G. Assanto, Mol.
Cryst. Liq. Cryst. {\bf421}, 197 (2004).

\bibitem{OL2005}M. Peccianti, C. Conti, and G. Assanto, \ol {\bf 30}, 415(2005).

\bibitem{peccianti-jnopm-o3}M. Peccianti, C. Conti, G. Assanto, A. De Luca, and C. Umeton, J. Nonl. Opt. Phys. Mat. {\bf12}, 525
(2003).

\bibitem{explain-1}When the optical beam is small enough, the
bourndary efffect of the NLC cell can be neglected, and the
configuration can approximate symmetry.

\bibitem{explain-2}The charateristic length $w_m$ defined in
Eq.~(\ref{wm}) is general. When $\theta_0=\pi/4$, $w_m$ defined by
us will be reduced to the charateristic value $R_c$ defined in
Refs.~\onlinecite{PRL2003} and \onlinecite{peccianti-jnopm-o3} .

\bibitem{cao-wlxb-05}Q. Guo, B. Luo, F. Yi, S. Chi, and Y. Xie,
Phys. Rev. E {\bf69}, 016602 (2004); N. Cao and Q. Guo, Acta Phys.
Sin. {\bf54}, 3688 (2005).

\bibitem{explain-4}The bias $V \approx
E_{RF}(d+\Delta)$,\cite{OL2005} where $d$ is the NLC cell thickness
and $\Delta$ is taken to approximately be $4\mu$m to consist with
experimnental data for our sample. $w_m(\theta_0)$ and $w_m(V)$ are
from Eq.~(\ref{wm}). The critical power is a numerical result of
Eqs.~(\ref{nlse0})-(\ref{pretilt})

\bibitem{explain-3}Rigorously speaking, except one of them, they are quasi-solitons
(breathers) rather than solitons because the fixed input power can
only equal exactly one of the critical powers for different biases,
but approximate the others in the case under consideration.


\end{thebibliography}
\end{document}